\documentclass[conference]{IEEEtran}

\usepackage{cite}
\usepackage{amsmath,amssymb,amsfonts}
\usepackage{graphicx}
\usepackage{textcomp}
\usepackage{xcolor}

\usepackage{svg}
\usepackage{graphicx}
\usepackage{orcidlink} 
\usepackage{multirow}%
\usepackage{amsthm}%
\usepackage{mathrsfs}%
\usepackage{manyfoot}%
\usepackage{booktabs}%
\usepackage{algorithm}%
\usepackage{algpseudocode}%
\usepackage{array}%

\usepackage{listings}%

\usepackage{tabularx}
\usepackage{float} 
\usepackage{tcolorbox}
\usepackage{mdframed}
\usepackage{enumitem} 
\usepackage{siunitx}
\usepackage{url}
\newtheorem{definition}{Definition}

\usepackage[caption=false,font=footnotesize]{subfig}

\def\BibTeX{{\rm B\kern-.05em{\sc i\kern-.025em b}\kern-.08em
    T\kern-.1667em\lower.7ex\hbox{E}\kern-.125emX}}
\begin{document}

\title{Theory-Driven Network Backbones\\for Bipartite Projections}

\author{\IEEEauthorblockN{Nikos Salamanos,\orcidlink{0000-0003-0643-3424}}
\IEEEauthorblockA{\textit{Data Science Section} \\
\textit{IT University of Copenhagen}\\
Copenhagen, Denmark \\
nsal@itu.dk}
\and
\IEEEauthorblockN{   
Paula Gheorghiade,\orcidlink{0000-0003-2335-2036}}
\IEEEauthorblockA{\textit{Social Resilience Lab}\\\textit{Department of History and Classical Studies} \\
\textit{Aarhus University}\\
Aarhus, Denmark\\
p.gh@cas.au.dk}
\and
\IEEEauthorblockN{
Tom Brughmans,\orcidlink{0000-0002-1589-7768}}
\IEEEauthorblockA{\textit{Social Resilience Lab}\\\textit{Department of History and Classical Studies}\\
\textit{Aarhus University}\\
Aarhus, Denmark\\
tb@cas.au.dk}
\and
\IEEEauthorblockN{Michele Coscia,\orcidlink{0000-0001-5984-5137}}
\IEEEauthorblockA{\textit{Data Science Section} \\
\textit{IT University of Copenhagen}\\
Copenhagen, Denmark \\
mcos@itu.dk}
}

\maketitle

\begin{abstract}
Collecting data, including network data, is subject to measurement error. Network backboning is the process of statistically validating each edge in a network to estimate the probability that the edge is born of noisy measurements. The current backboning literature provides approaches in two classes: structural and statistical. The classes differ in many respects, but they share one commonality: the test used to determine whether to filter out noisy edges is data-agnostic; it does not take into account what the network is representing. However, the nodes and edges of all non-synthetic networks correspond to specific real-world entities, whose dynamics determine how the edge-generating process operates. Such entities and their dynamics are often known and should not be ignored by the network backboning algorithm. In this paper, we propose the first in a new class of network backboning algorithms: a theory-driven backboning approach that allows us to encode known dynamics into the backboning process. The resulting backbone embeds the encoded theory and produces a network compatible with it, rather than one that obeys a data-agnostic network process, as in the structural and statistical backboning classes. We specifically focus on the case of projecting and backboning a bipartite structure. We do so by allowing users to specify an Agent-Based Model for the two types of agents (nodes) in the network and by performing Simulation-Based Inference to fit the observed bipartite structure and, as a result, reconstruct the true relationships between the two classes of agents. We test our framework both on synthetic data -- where we have perfect information about the edge-generating process -- and in a case study using archaeological data on amphorae retrieved from different archaeological sites. 
%
%
Our synthetic experiments demonstrate that theory-driven backboning outperforms established structural and statistical approaches in recovering latent network connections. In the archaeological case study, despite the imperfect fit of the generative model to the observed data, our approach identifies interpretable network patterns that highlight how explicit theoretical assumptions shape the extracted backbone.
\end{abstract}


\section{Introduction}
\label{sec:intro}

Measurement error is a crucial problem to solve in science and engineering and there are many statistical techniques to minimize its impact \cite{ott2010introduction}. Network data is affected by issues of measurement error as much as any other type of data. For instance, we could have type I errors where edges we observed are born out of noise. Type II errors involve missing edges that should be included in the network structure. Network error reconstruction tackles both error types at the same time \cite{wang2012measurement, newman2018network, peixoto2018reconstructing}, while link prediction focuses on type II errors \cite{martinez2016survey, kumar2020link, zhang2018link}.

Network backboning, the focus of our paper, specializes in addressing type I errors: to identify and eliminate edges included by mistake \cite{yassin2025backbone, yassin2025exploring, coscia2021atlas}. Backboning methods can currently be divided into two classes: structural and statistical. Structural backboning methods apply an algorithm on the topology of the network to identify which edges are playing an important structural role -- a classical example is to set a minimum threshold and eliminate all edges whose weight does not meet such threshold. Statistical backboning methods, on the other hand, create a null model with a different expected weight for each edge and then test whether the observed edge weight is significantly higher than its expectation.

In this paper, we propose to create a novel backboning approach that belongs to a third, novel class: theory-driven backbones. We want to address the common weakness that underpins both the structural and statistical backboning methods: their data-agnosticism. Structural backboning algorithms and statistical backboning null models do not take into account which real entities the nodes of a network represent. However, such information is often available to the analyst, together with a number of expectations about how such real entities connect in the real world. We propose a framework in which the analyst can embed this knowledge in a data-generating process, which then fits the observed data and produces a final network backbone.

We specifically target bipartite networks to produce backbones of their unipartite projections, therefore tackling in one step the problem of choosing a combination of a bipartite projection method and a backboning one \cite{coscia2019impact}. We address this problem through a framework that combines Agent-Based Models (ABMs) \cite{gilbert2019agent} and Simulation-Based Inference (SBI) \cite{cranmer2020:pnas:sbi}. The ABM provides an abstract view of the latent distribution process that generated the observed bipartite structure, with two types of agents representing the two node types in the bipartite network. SBI then asks which parameter values of the proposed ABM could plausibly have generated the observed bipartite network.

Because we have a generic ABM interface, the analyst can embed their theory of how entities in the real world connect in the ABM. The posterior simulations via SBI can then be used to embed the agents in an embedding space, where their true relations can be extracted. In this way, the inferred network would be grounded in an explicit edge generation process theory, rather than obtained directly from a data-agnostic projection and backboning of the observed bipartite structure. The framework is flexible and model-agnostic: different agent-based edge-generating processes can be embedded within the same SBI procedure.

We test our framework in two ways. First, we use synthetic data where we completely control the edge-generating process, and we can show directly the accuracy of the reconstructed backbones, as well as the superiority of embedding the right model over using model-agnostic methods such as structural or statistical backboning.
Second, we focus extensively on a real-world case study from archaeology to demonstrate practical usefulness. The case considers amphora types associated with production regions and their distribution across consumption sites in the Roman Empire. The aim of this empirical application is to demonstrate how ABM mechanisms can be specified, calibrated with SBI, and evaluated against observed amphora distributions. We compare simulated amphora-specific excavation-site distributions with the empirical distributions. The data selection reflects the period 1-200 CE based on the production dates of the amphora types.

In summary, the contributions of our paper are as follows:

\begin{itemize}
\item We introduce a new class of problems in the network backboning literature -- theory-driven backbones -- which significantly differ from any other backboning approach researched so far;
\item We provide the first framework to solve theory-driven backbones using a combination of ABM and SBI;
\item We develop the first theory-driven backboning method following our framework, a method designed specifically to solve the projection-backboning problem for bipartite networks;
\item We show the usefulness of theory-driven backbones, on synthetic data and a real-world case study using archaeological data.
\end{itemize}

\textbf{Code \& data availability:} Code and data are publicly available on GitHub at \url{https://github.com/nikos-salamanos/theory-driven-backbones}. 

\section{Related Work}
\label{sec:related-work}

In the introduction, we outline the network backboning literature as belonging fundamentally to two classes: structural and statistical. Here we discuss this distinction more in details, but we refer to recent surveys for a more in-depth treatment of the subject \cite{yassin2025backbone, yassin2025exploring}. We note that network backboning is such a common issue to face that there are software solutions for wider use outside the network science community \cite{neal2022backbone}.

In the structural class of network backboning, we have the double-stochastic approach based on the normalization of the rows and columns of the weighted adjacency matrix \cite{slater2009two}, the extraction of the High-Salience Skeleton based on short path trees \cite{grady2012robust}, and other \cite{gemmetto2017irreducible, zhang2018extracting, vsubelj2018convex, rajeh2022modularity}. None of these methods consider the statistical significance of an edge weight, but apply an algorithm on the topology of the network to identify which edges are playing an important structural role. In summary, the structural approach of backboning does not build a model giving expectations for edge weights, and is therefore agnostic about the edge-generating process.

In the statistical class of network backboning, the most used are the Disparity Filter (DF) \cite{serrano2009extracting}, the Noise-Corrected (NC) \cite{coscia2017backboning}, and others \cite{radicchi2011information, marcaccioli2019polya, siniscalco2026multilayer}. These approaches create a null model with an expected weight for each edge. They then test whether the observed edge weight is significantly higher than this expectation. The difference with the structural approach is that we now have a model for the edge weights. However, the commonality with the structural approach is the data agnosticism: the models proposed here are blind to the theoretical expectations we have about the real-world entities represented by the nodes. Each model in this class is generic and works with any network structure, regardless of what the structure represents. Our approach targets this shortcoming, allowing the backbone to use a specific theory for the edge-generating process, on which to base the backboning step.

We target bipartite networks specifically, and there are backboning methods developed for this case \cite{neal2014backbone, neal2021comparing, ramirez2025extracting}. However, such methods are still data-agnostic, besides adapting the null models to the specific connectivity patterns we expect in bipartite networks. Moreover, we aim to provide a unipartite backbone, essentially performing projection and backboning in one step. This is beneficial because there are also a number of network projection methods \cite{zhou2007bipartite, horvat2012one, melamed2014community, yildirim2014using, saracco2017inferring, cimini2022meta, gallo2025statistically}. Finding the right combination of a projection and backboning method that produces a meaningful unipartite network is a hard problem with a large search space and, often, daunting complexity \cite{coscia2019impact} -- there are no guidelines on how specific pairs of projection-backboning methods interact with one another. With our approach, we can bypass this problem altogether, as we perform both steps within the same, theory-driven framework.

\section{Methods}

\subsection{Problem Definitions}
In this section, we first introduce the general theory-driven backboning problem definition in general, then we move to the specific case we solve: starting from a bipartite view of the data and performing the backboning and projection steps at once.

We start by establishing some notation. Let $G = (V, E)$ be a directed graph, with node set $V$ and edge set $E$. We define each edge in $E$ to be a triple $(u, v, w)$, with $u,v \in V$ being the two nodes connected by an edge with weight $w \in \mathbb{Z}^{+}$ being a positive integer. Note that we can allow for any arbitrary positive real weight -- i.e., $w \in \mathbb{R}^{+}$. For simplicity -- and consistency with previous work -- we limit ourselves to discrete edge weights. 

In the case of a bipartite graph, we have $G = (V_1, V_2, E)$, with two nodes classes ($V_1$ and $V_2$). Edges are still $(u, v, w)$ triples as before, with the additional constraint that $u \in V_1$ and $v \in V_2$ without exception.

\subsubsection{Theory-Driven Backbones}
In the backboning literature, the assumption is that each edge is measured with a certain amount of error. Since there is uncertainty about whether a measured edge exists, we need to estimate the likelihood of an edge in a layer having an actual weight of zero -- i.e., it does not exist. This leads to the general backboning problem definition:

\begin{definition}
Given a directed graph $G = (V, E)$ with discrete count edge weights defined as $(u, v, w)$, define a function $f: G,(u, v, w) \rightarrow [0, 1]$ such that the output represents the likelihood of edge $(u, v, w)$ to have a weight $w > 0$.
\end{definition}

This fits all network backboning classes of solutions, which differ in how function $f$ is defined. Our novel problem definition can be amended so:

\begin{definition}
Given a directed graph $G = (V, E)$ with discrete count edge weights defined as $(u, v, w)$, and $M$ a data generating process for edge weights, define a function $f: G,(u, v, w),M \rightarrow [0, 1]$ such that the output represents the likelihood of edge $(u, v, w)$ to have a weight $w > 0$ according to the expectation defined by $M$.
\end{definition}

Note the crucial differences that set this problem definition apart from both the structural and statistical backboning classes. In the structural class, there is no model $M$. In the statistical class, model $M$ is not a parameter of $f$, but it is embedded in $f$ itself. In our theory-driven backbones, we explicitly request $M$ to be a parameter of $f$. The same $f$ can take different models $M$ and produce different backbones, while in the statistical literature having a different $M$ implies having a different $f$.

\subsubsection{Theory-Driven Backbones for Bipartite Networks}
In this paper, we want to work with a more specific problem definition: we want to start from a bipartite structure and produce the theory-driven backbones of its unipartite projections. We leave the solution of working directly with a unipartite weighted network for future work.

This means we have a different problem definition:

\begin{definition}
Given a bipartite $G = (V_1, V_2, E)$ with edge weights defined as $(u, v, w)$, and $M$ a data generating process for edge weights, define a function $f: G,(u_1, u_2), M, \phi \rightarrow \mathbb{R}$ such that $u_1,u_2 \in V_1$ and the output represents how strongly the observed projected similarity $\phi$ exceeds the corresponding projected similarities generated by $M$. 
\end{definition}

Higher values of f indicate stronger evidence for including the edge $(u_1,u_2)$ in the projected backbone. Note that in this definition, we only explicitly build one of the two possible projections of bipartite $G$, specifically the one containing the relationships between nodes in $V_1$. However, because the definition is symmetric -- it is still valid even if we swap $V_1$ with $V_2$ -- the same approach can be used to produce both projected backbones.

\subsection{Framework}

\begin{figure*}[t]
\centering
\includegraphics[width=0.99\textwidth]{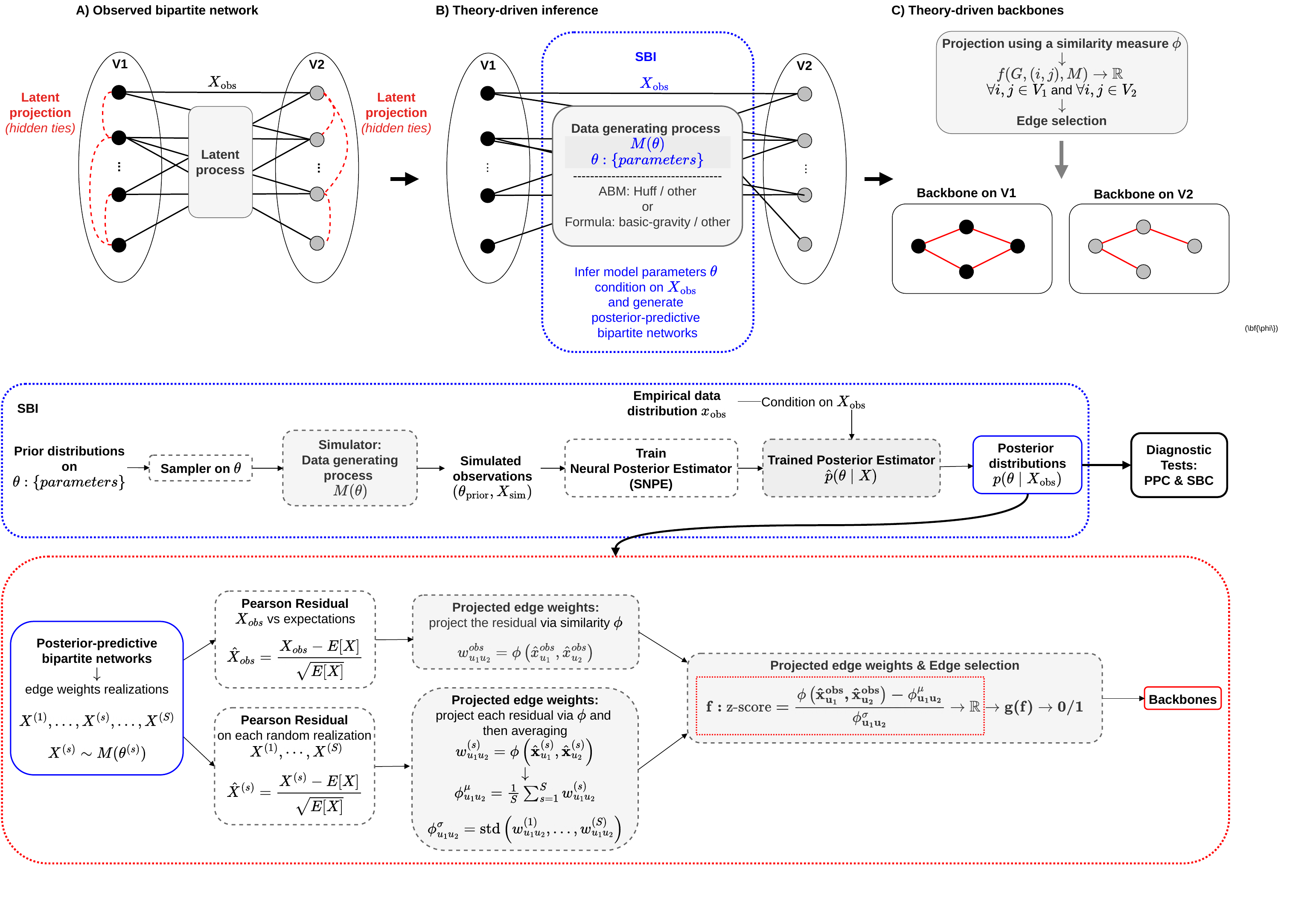}
\caption{ 
The conceptual framework of the TDBB method.
}
\label{fig:conceptual_technical_frameowrk}
\end{figure*}

Figure~\ref{fig:conceptual_technical_frameowrk} summarizes the conceptual and technical framework. Its upper panels illustrate the overall conceptual framework, the transition from an observed weighted bipartite network to theory-driven inference and, finally, to the projected backbones. The lower panels detail the technical workflow of SBI and the final backboning phase.

\subsubsection{From observations to theory-driven inference and backbones}
We consider an observed weighted bipartite network $G=(V_1,V_2,E)$, with $X_{\mathrm{obs}}$ edge weights. These observations are assumed to be the outcome of an underlying \emph{latent process} that cannot be observed directly (Figure~\ref{fig:conceptual_technical_frameowrk}A). In many empirical settings, the observed edge weights are therefore only an indirect and potentially noisy representation of this process.

We further assume that the observed bipartite edge weights reflect a systematic signature of the latent relationships between the two node classes, $V_1$ and $V_2$. These latent relationships are not directly observed, and our objective is to recover the part of their structure that is supported by the observed bipartite network.

In order to represent the latent process, we specify a theory-driven data-generating model $M(\theta)$, where $\theta$ denotes the model parameters (Figure~\ref{fig:conceptual_technical_frameowrk}B). The model may range from a simple parametric formulation, such as a basic gravity model, to a complex agent-based mechanism. Then we deploy Simulation-Based Inference (SBI) to condition the model $M$ on $X_{\mathrm{obs}}$ and infer the posterior distribution $p(\theta\mid X_{\mathrm{obs}})$. Finally, we sample from the posterior thousands of parameter $\theta$ samples to generate posterior-predictive bipartite networks, which provide us the model-based input for the final projection and backboning phase (Figure~\ref{fig:conceptual_technical_frameowrk}C).

\paragraph{\textbf{Simulation-Based Inference (SBI)}}
 The role of model $M(\theta)$ (either ABM or formula) is to emulate, at an abstract level, the latent process that generated the $X_{\mathrm{obs}}$. A Simulation-Based Inference (SBI) method is used to calibrate these parameters $\theta$ against the empirical observations (see the middle panel of Figure~\ref{fig:conceptual_technical_frameowrk}). In this way, the noise in the empirical data is partially handled through Bayesian inference. We employ Sequential Neural Posterior Estimation (SNPE)~\cite{papamakarios2016fast:snpe} using the Automatic Posterior Transformation (APT) algorithm~\cite{greenberg2019automatic:sbi:implementation:snpe}, as implemented in the \texttt{sbi} Python library~\cite{BoeltsDeistler_sbi_2025}. SNPE assumes access to a forward simulator that maps model parameters $\theta$ to synthetic observations $X_{\mathrm{sim}}$, and learns an approximation to the posterior distribution $p(\theta \mid X)$ without requiring an explicit likelihood function. 
 
After training, the neural posterior estimator is conditioned on the empirical observation $X_{\mathrm{obs}}$, and posterior samples are drawn from the resulting approximation to $p(\theta \mid X_{\mathrm{obs}})$. These samples provide an empirical representation of the posterior distribution over the $\theta$ parameters.

\paragraph{\textbf{Diagnostic Tests}} 
There are two main diagnostic procedures under the SBI framework: the \textbf{posterior predictive checks (PPC)}, which evaluate the empirical adequacy of the calibrated ABM with respect to the observed empirical distributions, and \textbf{simulation-based calibration (SBC)}, which evaluates the reliability of the inference procedure itself. If SBC indicates good calibration but posterior predictive checks fail, then the problem is more likely to lie in the model specification rather than in the neural posterior estimator.

\noindent 
\textbf{PPC} asks whether the observed $X_{\mathrm{obs}}$ are plausible within the model's predicted range, given posterior uncertainty. The output of SBI is not a single best parameter vector, but a posterior distribution $p(\theta \mid X_{\mathrm{obs}})$ that represents the range of parameter values compatible with the empirical data. To perform PPC, we sample parameter values $\theta^{(s)}$ from this posterior and pass them through the model M simulator to generate posterior predictive observations $X_{\mathrm{ppc}}^{(s)}=f(\theta^{(s)})$. The collection of simulated values defines a posterior predictive distribution, which is compared with the corresponding empirical probability in $X_{\mathrm{obs}}$. 

\noindent
\textbf{SBC} asks whether the trained posterior estimator is statistically calibrated under the assumed prior and simulator. The procedure is therefore applied to synthetically generated data, for which the true generating parameters are known. We first sample a parameter vector $\theta_{\mathrm{true}}$ from the prior and pass it through the ABM simulator to generate a synthetic observation $X_{\mathrm{sim}}$. The trained neural posterior estimator is then conditioned on $X_{\mathrm{sim}}$ to obtain posterior samples from $\hat{p}(\theta \mid X_{\mathrm{sim}})$. Since $\theta_{\mathrm{true}}$ is known, we can evaluate whether it is consistent with the inferred posterior distribution, for example, through rank-based diagnostics. 

\paragraph{\textbf{Projection \& Backboning}} 
Once we have fitted the $\theta$ parameters, we can compare $X_{\mathrm{obs}}$ against the expected edge weights $E[X]$ generated by the fitted model $M$ (see the lower panel of Figure~\ref{fig:conceptual_technical_frameowrk}). We do so by calculating the Pearson residuals against these expectations:
$\hat{X}_{\mathrm{obs}} = (X_{\mathrm{obs}} - E[X]) / \sqrt{E[X]}$. 
Then, we project these residual bipartite edge weights into a unipartite $G_1$ projection, using an arbitrary similarity measure $\phi$ -- co-occurrence counts, cosine, or Pearson similarity. Hence, the observed projected edge weight is
$w^{\mathrm{obs}}_{u_1u_2} = \phi\left(\hat{x}^{u_1}_{\mathrm{obs}},\hat{x}^{u_2}_{\mathrm{obs}}\right)$.

This is not yet the theory-driven score required by Definition 3, because random realizations of $M$ can still produce projected similarities that surpass the observed one. Therefore, we generate a number of posterior-predictive realizations $X^{(1)}, \ldots, X^{(s)}, \ldots, X^{(S)}$ -- generated by $M$, i.e. $X^{(s)} \sim M(\theta^{(s)})$ and dependent on $|V_1|$ and $|V_2|$ -- and we calculate their Pearson residuals against the same expectation,
$\hat{X}^{(s)} = (X^{(s)} - E[X]) / \sqrt{E[X]}$. 
We project each residual realization using the same similarity measure $\phi$, obtaining
$w^{(s)}_{u_1u_2} = \phi\left(\hat{x}^{(s)}_{u_1},\hat{x}^{(s)}_{u_2}\right)$,
and we calculate the mean $\phi^\mu_{u_1u_2}$ and standard deviation $\phi^\sigma_{u_1u_2}$ for each projected edge.

We can now formulate the output of our desired function $f$, which quantifies how strongly the observed projected similarity exceeds the corresponding projected similarities generated by the fitted model $M$:

$$
f:G,(u_1,u_2),M,\phi \rightarrow
\frac{
\phi\left(\hat{x}^{u_1}_{\mathrm{obs}},\hat{x}^{u_2}_{\mathrm{obs}}\right)
-\phi^\mu_{u_1u_2}
}{
\phi^\sigma_{u_1u_2}
}
\in \mathbb{R},
$$

with $\phi^\mu_{u_1u_2}$ and $\phi^\sigma_{u_1u_2}$ calculated from the posterior-predictive realizations generated by $M$, and $\hat{X}_{\mathrm{obs}}$ obtained by comparing $X_{\mathrm{obs}}$ with $E[X]$. Higher values of $f$ indicate stronger evidence for including the edge $(u_1,u_2)$ in the projected backbone. Finally, a backbone can be obtained by applying a binary edge-selection rule $g(f) \rightarrow 0/1$ to the scores returned by $f$.

\section{Validation}

\subsection{Synthetic Data}
To properly validate our approach we need to generate synthetic data that follows a clear edge generating process and exactly matches the assumptions of the framework. We do so in three steps, following our assumptions: (i) use a global model $M$ that affects all nodes equally to create the baseline bipartite edge weights $B_M$; (ii) create the ground truth real projected edges -- $G_1$ and $G_2$ for both possible projections of $B$ -- which operate outside the global model's control; (iii) use a function $f$ to increase/decrease the global model's bipartite edge weights by using proximities between the nodes in the ground truth projections, resulting in the final $B = B_M \cdot h\left(G_1, G_2\right)$. 

By following this procedure, once we apply our framework with a model $M' \sim M$ to $B$ we take out from $B$ a $B_{M'}$, leaving out only an approximation of the information contained in $G_1$ and $G_2$, which we can use to reconstruct them. We now detail how we implement each step.

\subsubsection{Synthetic Global Model}
Here we use a simple gravity model. For each node in $V_1$ and $V_2$ we draw a size $\sigma$ from a random lognormal distribution. Then, for each pair of $V_1$ and $V_2$ nodes, we draw a distance value $\delta$ from a censored normal distribution: first we draw costs with average $0$ and standard deviation of $1$, then we take the absolute values and add a $0.5$ cost constant to avoid exceedingly low costs. In the end, edge weights in $B_M$ are produced as follows:

$$ B_M(v_1,v_2) = \dfrac{\sigma_{v_1} \sigma_{v_2}}{\delta_{v_1,v_2}}. $$

\subsubsection{Synthetic Ground Truth}
The real ground truth edges in both projections $G_1$ and $G_2$ use the exact same algorithm with slightly different parametrization to create a small asymmetry between nodes in different projections.

We extract edges using a Stochastic Block Model (SBM) \cite{holland1983stochastic}. We first create the community set $C$, and each community $c \in C$ has a random size drawn from a power law distribution with exponent $\alpha$ (a parameter of the generator), creating one large community and many small ones around it. We ensure each community has at least two nodes by redrawing the community size vector until this condition is satisfied.

Then we create the probabilities of nodes to connect within their community ($p_{in}$) and outside their community ($p_{out}$) so that we can guarantee an average degree within the community that is higher than the average number of edges a node sends outside the community:

$$ p^c_{in} = \min \left( \dfrac{2k}{|c| - 1}, 1 \right), \text{ and } p_{out} = \dfrac{\mu}{\sum \limits_{c\in C} |c|},$$

where $|c|$ is the size of the community $c$ in number of nodes. Additionally, $k$ is the average degree within a community and $\mu$ is the mixing parameter regulating how many inter-community edges there are. Both $k$ and $\mu$ are parameters that can be set, note that $\mu$ is the same for all communities.

By knowing $p_{in}$ and $p_{out}$ we have fully specified an SBM and we can draw a random network from it. For one projection we set $|C| = 3$ (number of communities), $\alpha = 1$, $max(|c|) = 30$ (the maximum community size, to scale the size of the projection), $\mu = 0.8$, $k = 3$. For the other projection we instead set $|C| = 4$ (more communities), $\alpha = 2$ (communities tend to have smaller size variation), $max(|c|) = 20$ (communities are smaller), $\mu = 0.9$ (communities are less well separated), $k = 3$. 

We ensure that both projections are a single connected component and contain at least $30$ nodes, redrawing a sample if either condition is not satisfied.

\subsubsection{Synthetic Observed Bipartite}
To create the final observed synthetic $B$ we need to augment the edges weights of $B_{M}$ with the topological information from the latent projections $G_1$ and $G_2$. We do so by increasing the edge weights of $B_{M}$ if the two connected nodes are also connected to ``nearby'' nodes in the two projections.

To quantify ``nearby'' we exploit the normalized effective resistance matrices $\Omega_1$ and $\Omega_2$ of $G_1$ and $G_2$, which give us a proper metric over a graph's topology in a way shortest paths do not \cite{devriendt2022effective, coscia2021atlas}. Given an undirected graph $G$, $\Omega_{ij} = \Gamma_{ii} + \Gamma_{jj} - 2\Gamma_{ij},$ with $\Gamma = \left( L + 1 / |V| \right)^\dagger$ -- i.e. the pseudoinverse of the normalized Laplacian. Because $\Omega$ is a distance, we transform it into a similarity via an exponentiation: $\tilde{\Omega} = e^{-\Omega/\tau},$ with $\tau = 0.1$ being a temperature parameter to increase contrast.

To end up with a $|V_1| \times |V_2|$ similarity matrix for the bipartite network, we multiply the smoothed versions of the two normalized effective resistance similarities ($\tilde{\Omega}_1 \cdot \tilde{\Omega}_2'$). We can now multiply this result to $B_{M}$:

$$ B = B_M \cdot e^{s (\tilde{\Omega}_1 \cdot \tilde{\Omega}_2')},$$

with $s = 2$ a strength parameter that we can use to make the latent structure of $G_1$ and $G_2$ dominate over $B_M$. With $s = 0$, only $B_M$ remains and there is no latent structure to discover by modeling it; with high $s$ values the latent structure dominates over the global model.

\subsection{Results}
We generate $100$ independent instantiations of $B_{M'}$ with its corresponding $G_1$ and $G_2$ latent projections. We apply a number of backboning algorithms. We evaluate each backboning algorithm with its ability of recovering the original edges of the latent $G_1$ and $G_2$ projections. We use both ROC and Precision-Recall curves, and summarize them with their respective AUC scores, together with the 95\% confidence intervals to establish the significance of the differences across different runs.

We pick naive thresholding as the basic structural approach. For the statistical approaches we consider Disparity Filter (DF) \cite{serrano2009extracting} -- the most used backboning algorithm --, and Noise-Corrected (NC) \cite{coscia2017backboning} -- which we expect to perform well because it has a proper edge-level model which can be considered as a theory-agnostic backboning technique.

We run two versions of our theory-driven backboning framework. The first (Gravity) uses as its ABM a gravity model that corresponds to the $M$ global model we use for $B_M$. The second (Random) uses a uniform random ABM where the edges of $B$ are modeled completely at random. We do this to show the performance difference in having the right model $M$ against not having it.

\begin{figure}[t]
\centering
\includegraphics[width=0.49\columnwidth]{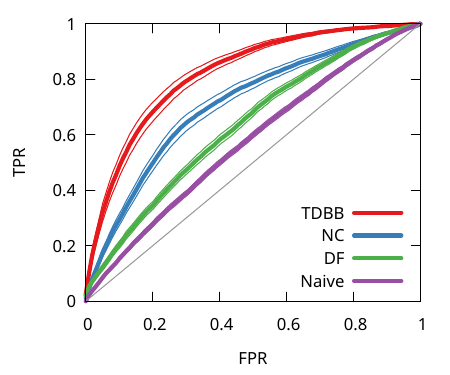}
\includegraphics[width=0.49\columnwidth]{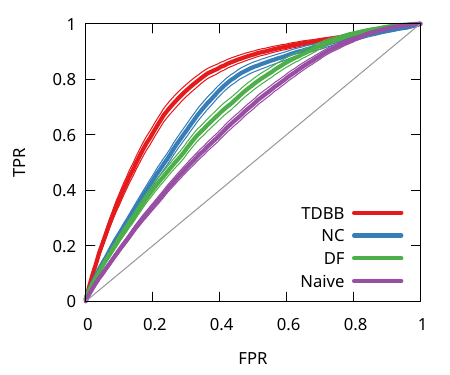}
\caption{The ROC curves for predicting the latent edges in $G_1$ (left) and $G_2$ (right) for the various backboning methods (line color). Thick line for average, thin lines for 95\% confidence intervals.}
\label{fig:synth-gravity-roc}
\end{figure}

\begin{figure}[t]
\centering
\includegraphics[width=0.49\columnwidth]{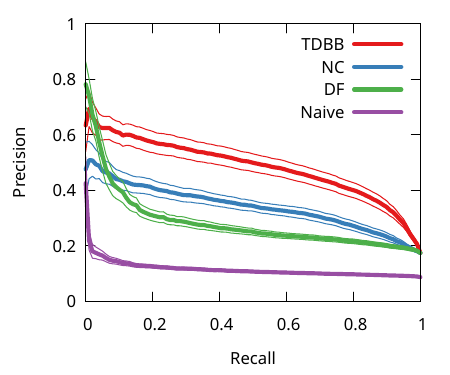}
\includegraphics[width=0.49\columnwidth]{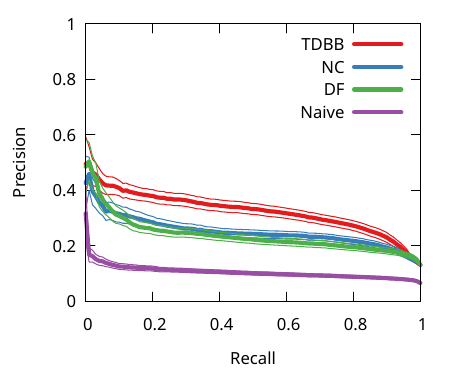}
\caption{The Precision (y axis) - Recall (x axis) curves for predicting the latent edges in $G_1$ (left) and $G_2$ (right) for the various backboning methods (line color). Thick line for average, thin lines for 95\% confidence intervals.}
\label{fig:synth-gravity-pr}
\end{figure}

First, Figures \ref{fig:synth-gravity-roc} and \ref{fig:synth-gravity-pr} report the ROC and Precision-Recall curves of our Theory-Driven BackBone (TDBB) using the correct gravity model, against the non-theory driven backbones. We have two subplots not to mix the two different node types -- we predict the edges in the latent $G_1$ and $G_2$ independently.

We can see that, in all cases, there is a clear performance hierarchy: Naive $<$ Disparity-Filter $<$ Noise-Corrected $<$ TDBB. This matches well with expectation: Naive does not do any edge modeling and performs the worst. Noise-Corrected has an edge-level model against Disparity Filter which has a naive node-level model, and that is how NC can easily outperform DF. Finally, TDBB has the most accurate model of them all and it dominates performance.

Note how performances are closer in $G_2$, which is the projection with fewer communities dominated by the largest ones, where the topological information is less clear. It is therefore natural that knowing the correct model is less useful for this specific projection.

\begin{figure}
\centering
\includegraphics[width=0.49\columnwidth]{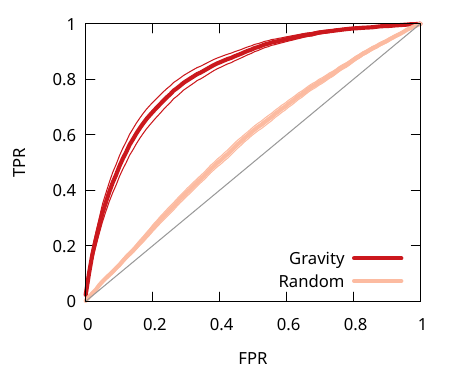}
\includegraphics[width=0.49\columnwidth]{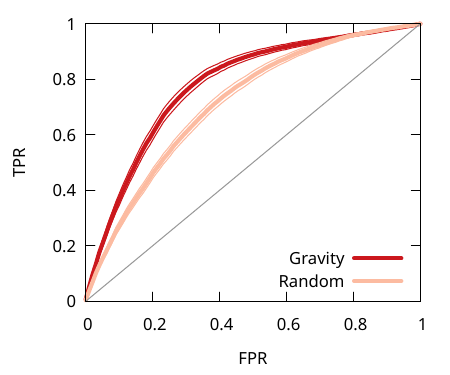}
\caption{Same as Figure \ref{fig:synth-gravity-roc}, comparing the Gravity TDBB (dark red) against the Random TDBB (bright red).}
\label{fig:synth-gravity-roc-rnd}
\end{figure}

\begin{figure}
\centering
\includegraphics[width=0.49\columnwidth]{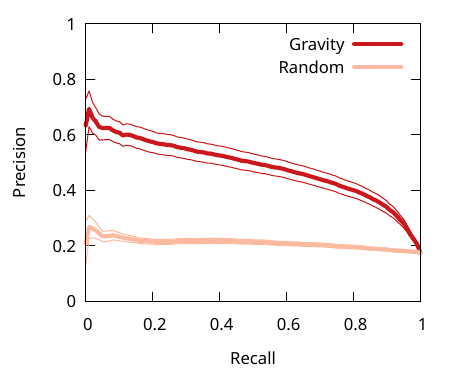}
\includegraphics[width=0.49\columnwidth]{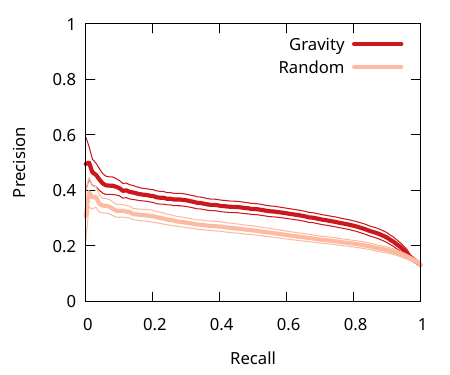}
\caption{Same as Figure \ref{fig:synth-gravity-pr}, comparing the Gravity TDBB (dark red) against the Random TDBB (bright red).}
\label{fig:synth-gravity-pr-rnd}
\end{figure}

Figures \ref{fig:synth-gravity-roc-rnd} and \ref{fig:synth-gravity-pr-rnd} report the ROC and Precision-Recall curves, this time comparing the gravity TDBB against the random TDBB. As expected, knowing the right gravity model increases the quality when recovering the latent projections.

However, also in this case the performance margin is reduced for the $G_2$ projection. This can be explained by the fact that $G_2$'s structure is closer to the one of a random network, and our framework is able to partially recover it using a random null model. In fact, this is a strength of the TDBB approach: one does not need to have a perfect model to recover good backbones. An accurate model will always be preferred, but an inaccurate model can still lead to reasonable backbones.

\begin{table}[t]
\label{tab:all-aucs}
\centering
\caption{The AUC values for the curves in Figures \ref{fig:synth-gravity-roc} to \ref{fig:synth-gravity-pr-rnd}, with the 95\% confidence intervals.}
\begin{tabular}{lll|c}
Curve & Proj & Method & AUC\\
\hline
\multirow{10}{*}{PR} & \multirow{5}{*}{$G_1$} & Naive & $0.115 \pm 0.006$ \\
 &  & DF & $0.282 \pm 0.017$ \\
 &  & NC & $0.341 \pm 0.023$ \\
 &  & TDBB-Random & $0.210 \pm 0.010$ \\
 &  & TDBB-Gravity & $\textbf{0.485} \pm 0.032$ \\
\cline{2-4}
 & \multirow{5}{*}{$G_2$} & Naive & $0.105 \pm 0.008$ \\
 &  & DF & $0.237 \pm 0.018$ \\
 &  & NC & $0.251 \pm 0.018$ \\
 &  & TDBB-Random & $0.255 \pm 0.017$ \\
 &  & TDBB-Gravity & $\textbf{0.325} \pm 0.022$ \\
\hline
\multirow{10}{*}{ROC} & \multirow{5}{*}{$G_1$} & Naive & $0.572 \pm 0.008$ \\
 &  & DF & $0.632 \pm 0.012$ \\
 &  & NC & $0.709 \pm 0.017$ \\
 &  & TDBB-Random & $0.571 \pm 0.010$ \\
 &  & TDBB-Gravity & $\textbf{0.819} \pm 0.016$ \\
\cline{2-4}
 & \multirow{5}{*}{$G_2$} & Naive & $0.645 \pm 0.011$ \\
 &  & DF & $0.689 \pm 0.014$ \\
 &  & NC & $0.715 \pm 0.014$ \\
 &  & TDBB-Random & $0.715 \pm 0.012$ \\
 &  & TDBB-Gravity & $\textbf{0.780} \pm 0.013$ \\
\end{tabular}
\end{table}

Table \ref{tab:all-aucs} confirms that the closer margin between TDBB-Gravity and TDBB-Random in $G_2$ is not only due to TDBB-Gravity performing worse. In a structure- and information-rich latent case ($G_1$) TDBB-Random performs poorly, on par with naive backboning. However, in a less crisp case ($G_2$) TDBB-Random is on par with Noise-Corrected, while the more accurate TDBB-Gravity dominates in all cases.

\section{Case Study}
\label{sec:case_study}

Because of the flexibility of the ABM+SBI framework, we can apply theory-driven backbones to networks representing any real-world phenomenon: users reading stories on social media, customers purchasing products in (online) retail, music bands sharing musicians, etc. For our case study, we decide to focus on a single application to show the method's applicability in depth, rather than applying it broadly across many applications -- which is possible if proper models are designed. We choose the case of archaeology because it is one of the toughest fields when it comes to data availability. In archaeology, we need to work with what we find, which is, by definition, a sample highly affected by preservation and selection biases. Preservation of ancient remains is highly variable due to natural and cultural disruptive effects (e.g., erosion, decomposition, and urban development); excavation activities are not uniform but patchy, based on construction activities and heritage management practices; and recording and publication practices are highly unstandardized. Therefore, the field allows to specify many different theories, as well as requiring to model the data loss process itself. 

For our case study, we choose to explore the circulation of foodstuffs (olive oil, wine, fish sauce) in the Roman Empire, which can be traced by analyzing ceramic amphora containers excavated at sites. These sites could have been cities, military settlements, or villas, which can be interpreted as destinations or consumption areas for amphora-borne foodstuffs. The empirical data arise from a complex process that may include various forms of noise and biases that have been introduced at different historical phases or during the excavation phase itself. Therefore, the amphora distribution is merely an archaeological trace of a hidden, multiphase latent process and should not be interpreted as direct evidence of interaction intensity or network ties between archaeological sites. The true unobserved process is that producers made amphora-borne goods; goods moved along routes and through intermediaries; the consumption sites had different demands; some places were more reachable on the transport system; some amphorae survived, and some were excavated and published, while the rest were lost.

What we actually observe is the archaeological evidence discovered and documented at the excavation sites. These are the amphora frequencies: Site 1 has many amphorae of type X from production area A; Site 2 has amphorae from both production areas A and B; and so on. But even these frequencies are the archaeological outcomes of research, excavation, and recording, each of which carries its own biases. For example, the archaeologists may have followed different methods for the final amphorae enumeration, such as Minimum Number of Individuals (MNI), Maximum Number of Individuals (MAX), and diagnostic Rim, Base, and Handles count (RBH)~\cite{franconi-2023-paper}. In other words, the observed data are noisy and sparse, and cannot fully represent the outcome of the latent distribution process. 

We address this problem through the TDBB framework. The spatial-choice provides the data-generating process $M(\theta)$, an abstract view of the latent distribution process that generated the observed amphora frequencies. SBI then asks which parameter values of the ABM could plausibly have generated the empirical amphora pattern. 

\subsection{The Data}
\label{subsec:raad-data}

\begin{table}[t]
\centering
\label{tab:archaeological_data}
\caption{Summary of the archaeological dataset.}
\small
\begin{tabular}{lr}
\hline
\textbf{Dataset characteristic} & \textbf{Count} \\
\hline
Consumption sites & 79 \\
Amphora types & 131 \\
Amphora-borne goods & 3 \\
Major production regions & 5 \\
Production origins (major region, subregion) & 24 \\
Distinct subregion labels & 20 \\
(major region, good) combinations & 12 \\
Possible amphora–site pairs & 10,349 \\
Observed (nonzero) edges & 891 \\
Total recorded amphora frequency & 27,531 \\
\hline
\end{tabular}
\end{table}

The archaeological data are obtained from the RAAD dataset\footnote{https://github.com/Tom-Brughmans/RAAD}~\cite{franconi-2023-paper}, selecting a subset of amphora types with known production origins, carrying wine, olive oil, or fish sauce, and production dates in the period 1--200 CE. The dataset contains five major production regions and three amphora-borne goods, with 12 region--good combinations represented in the observations. Table~\ref{tab:archaeological_data} summarizes the dataset and the resulting weighted bipartite network connecting amphora types to consumption sites. Unobserved edges do not necessarily imply historical absences.

Population estimates for consumption sites were compiled using a hierarchical procedure based firstly on Hanson’s dataset,  secondly on Wilson’s estimates, thirdly on online gazetteers, and fourthly on published literature~\cite{wilson-population-size}. For sites for which no population estimates could be established using these sources in this study, the mean population size of all other settlements (7126) is used. See the GitHub repository\footnote{https://github.com/nikos-salamanos/theory-driven-backbones} for RAAD data selection and population estimates.

For transport cost estimates, production origins and consumption sites were mapped to representative nodes in the ORBIS network\footnote{https://orbis.stanford.edu/}. These nodes serve as geographical proxies rather than precise locations of amphora production. Transport times, expressed in days, were calculated for the cheapest summer journeys using standardized transport settings.

Each amphora type $a$ is associated with an amphora-borne good (wine, olive oil, or fish sauce) and a production origin $o(a)$,
which is defined by two geographical locations (major region, subregion). 

The empirical observations $X_{\mathrm{obs}}$ consist of aggregated amphora counts by consumption site and amphora type. Let $freq_{i,a}$ denote the observed count of amphora type $a$ at consumption site $i$. Then we normalize across the consumption sites to obtain the empirical distribution of sites $i\in S$ conditional on amphora types:
$\hat{p}(i \mid a) = \displaystyle\frac{freq_{i,a}}{\sum_{i'\in S} freq_{i',a}}$

The observed data vector $X_{\mathrm{obs}}$ is then formed by concatenating these normalized site distributions across the amphora types. After training, the neural posterior estimator is conditioned on the empirical observation $X_{\mathrm{obs}}$, and posterior samples are drawn from the resulting approximation to $p(\theta \mid X_{\mathrm{obs}})$. These samples provide an empirical representation of the posterior distribution over the ABM parameters.

\begin{figure}
\centering
\includegraphics[width=0.9\columnwidth]{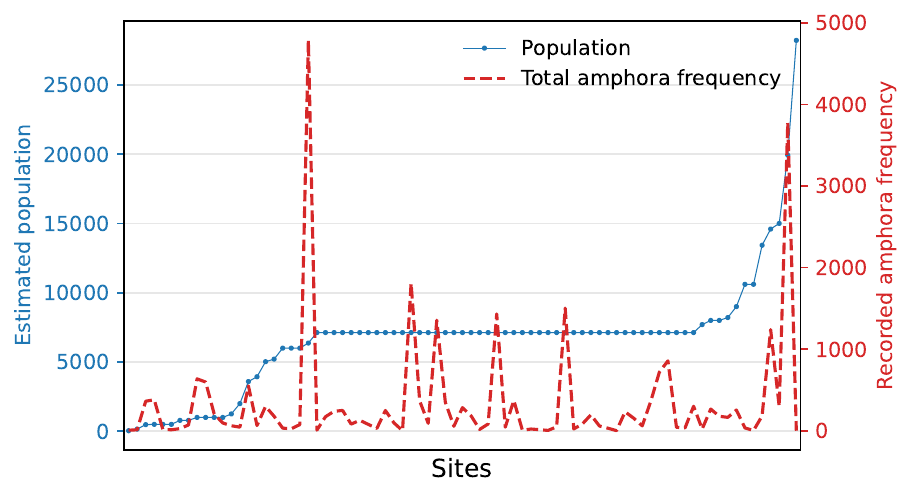}\\
\caption{Estimated population and total recorded amphora frequency across the 79 consumption sites, sorted by increasing population.}
\label{fig:population_vs_frequencies}
\end{figure}

\begin{figure*}[t]
\centering
\includegraphics[width=0.32\textwidth, trim={10cm 3cm 6cm 1.5cm}, clip]
{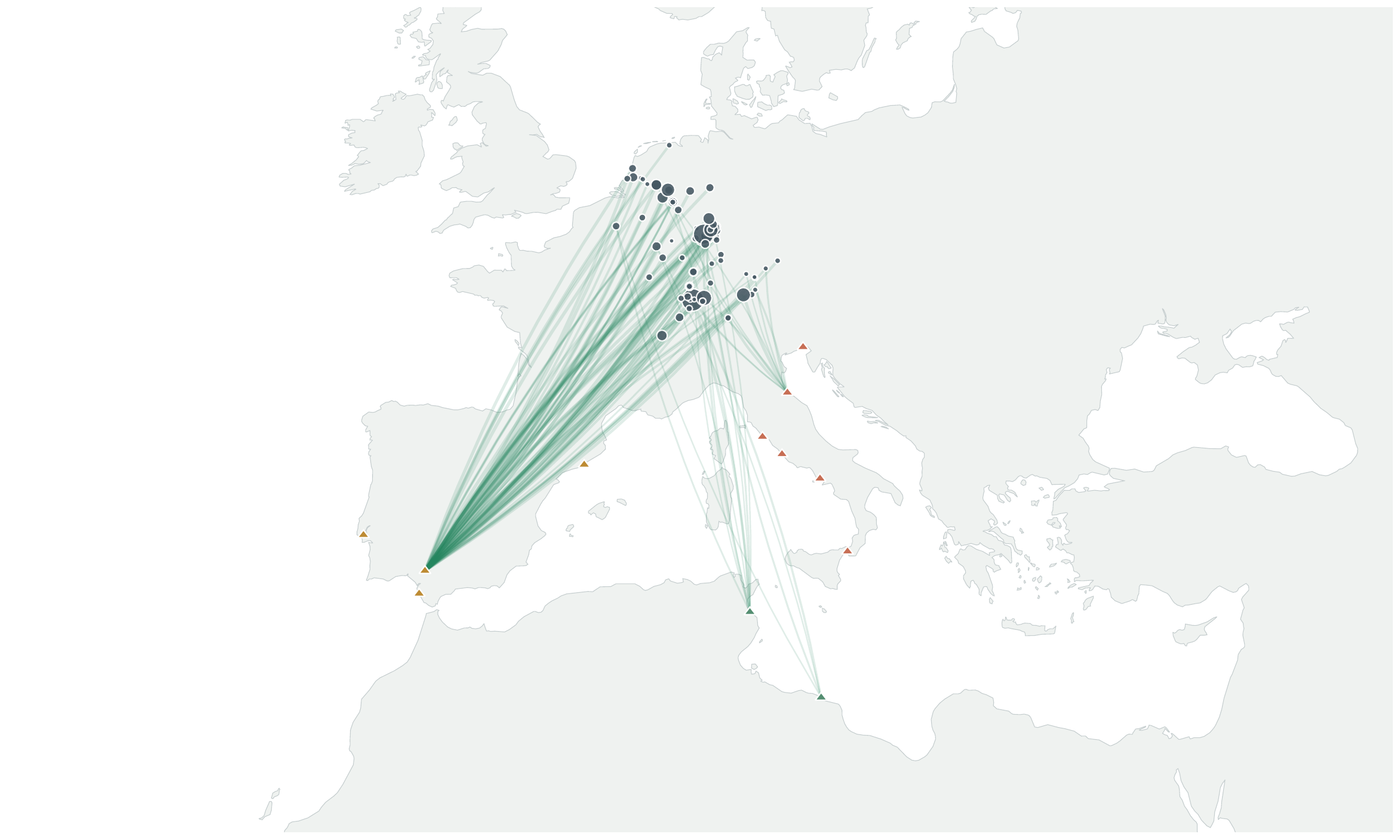}
\includegraphics[width=0.32\textwidth, trim={10cm 3cm 6cm 1.5cm}, clip]
{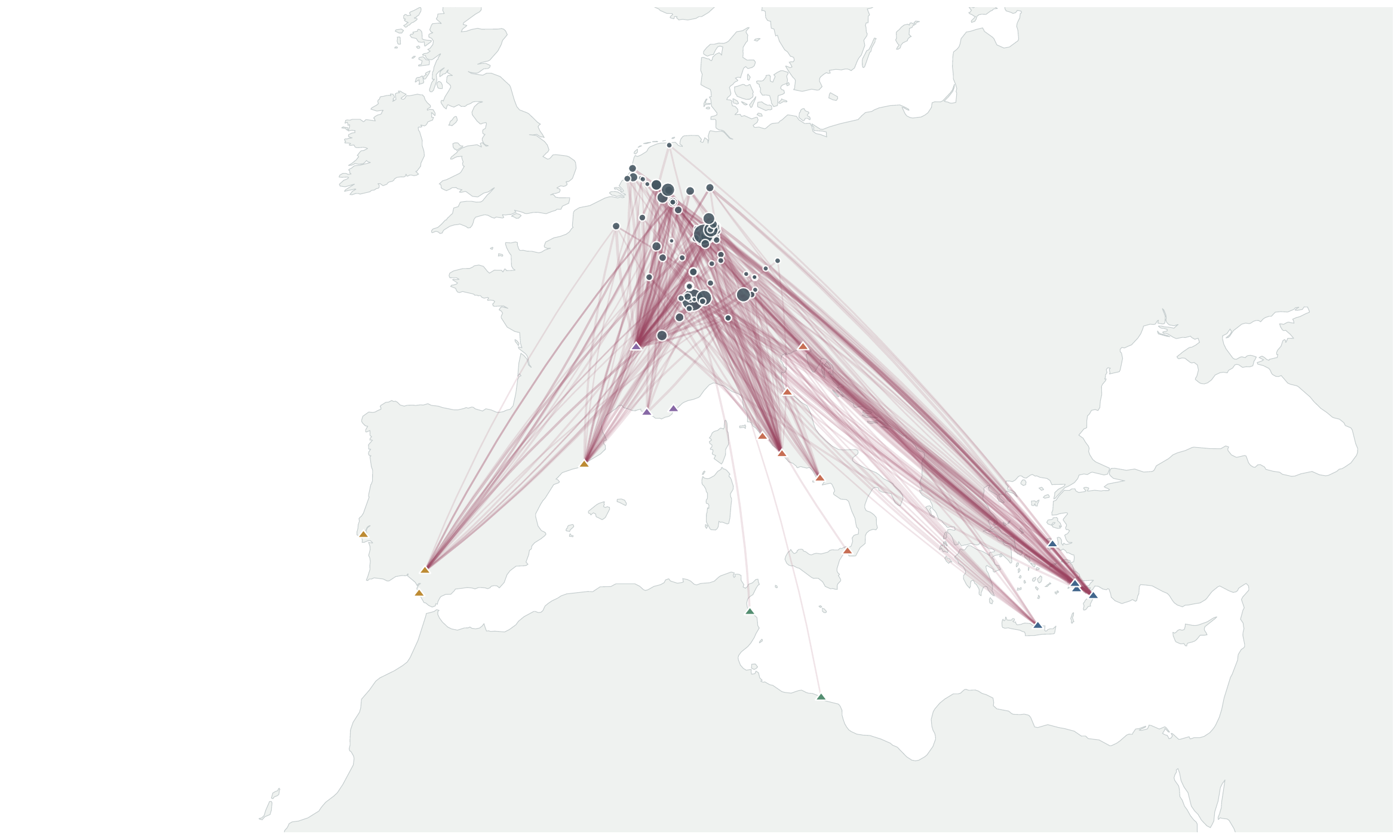}
\includegraphics[width=0.32\textwidth, trim={10cm 3cm 6cm 1.5cm}, clip]
{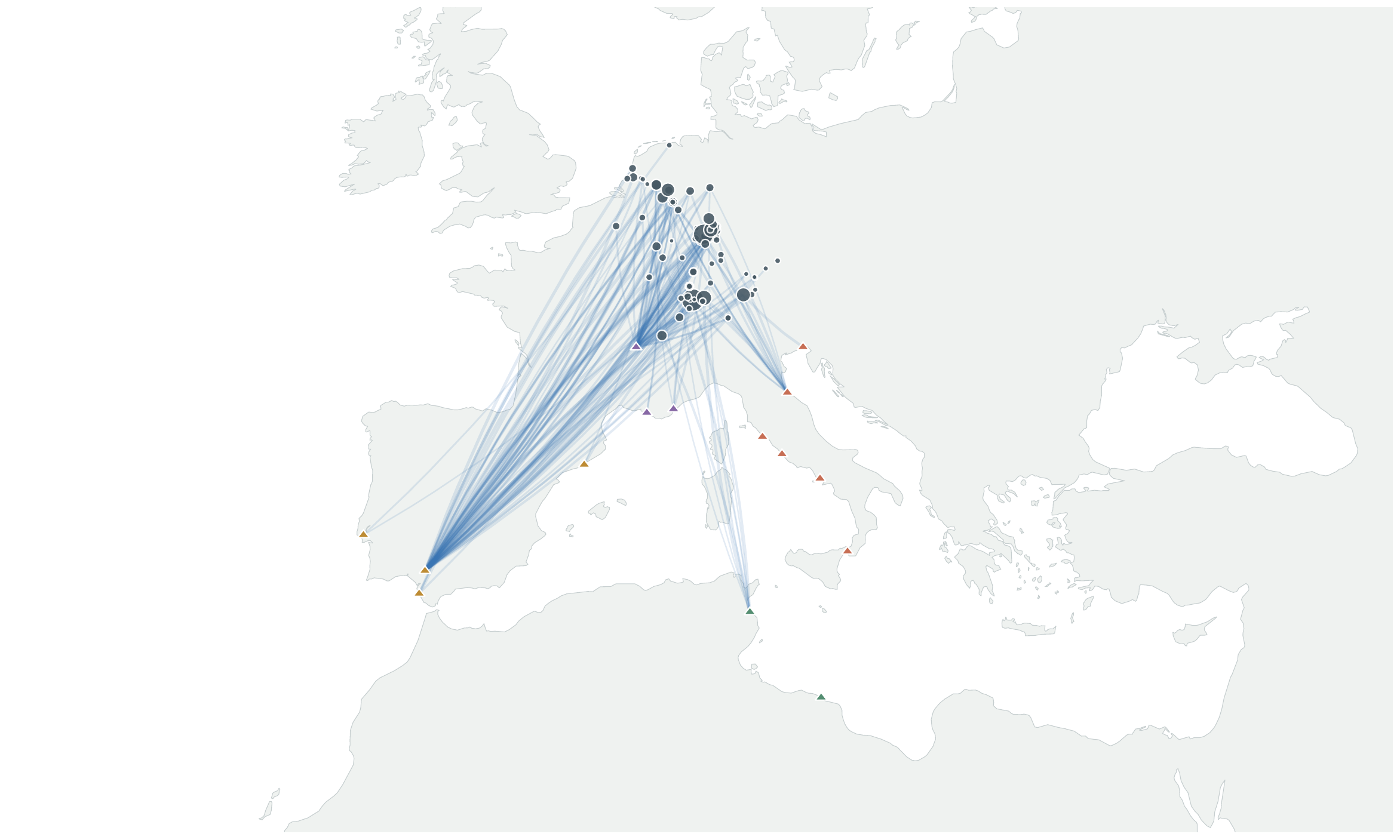}
\caption{
Geographical distribution of recorded amphorae carrying olive oil (left), wine (center), and fish sauce (right). Lines connect the estimated production-origin locations to consumption sites where amphorae were recorded. Line thickness reflects recorded amphora frequencies. These connections represent archaeological associations rather than reconstructed historical transport routes.
}
\label{fig:amphora_map}
\end{figure*}

\textbf{Experimental Rationale:}
The empirical investigation serves as a thought experiment motivated by two hypotheses about the circulation of amphora-borne goods in the Roman Empire: (i) consumption sites with larger populations generate greater demand and, all else being equal, receive larger quantities of amphorae; (ii) the allocation of amphora types is positively influenced by production-region attractiveness for
specific amphora-borne goods and negatively by transport costs.
However, the archaeological observations reveal little association between estimated site populations and recorded amphora frequencies (Figure~\ref{fig:population_vs_frequencies}), while showing pronounced geographical specialization by amphora-borne good (Figure~\ref{fig:amphora_map}). These patterns may partly reflect heterogeneous excavation intensity, preservation, and recording practices rather than historical demand alone.

Therefore, in the next section, we approach the Huff model as a theory-driven empirical thought experiment, rather than an attempt to reconstruct historical trade relationships directly. By calibrating this simplified spatial-choice mechanism against archaeological observations, we examine its empirical adequacy and explore the residual site--site similarities that remain unexplained by its assumptions. Such similarities are treated as exploratory patterns rather than evidence of historical connections, unless supported by independent archaeological validation.

\subsection{\textbf{Spatial-choice allocation models}}
\label{subsec:spatial_ABM}
\begin{algorithm}[t]
\caption{Spatial-choice agent-based allocation model}
\label{alg:spatial_ABM}
\small
\begin{algorithmic}[1]
\algrenewcommand\algorithmicrequire{\textbf{INPUT:}}
\algrenewcommand\algorithmicensure{\textbf{OUTPUT:}}

\Require\\
A set of nodes $V_1$. Each node $i \in V_1$ has $N_i$ demand agents.\\
A set of nodes $V_2$ representing the available options.\\
A \textbf{transaction-cost} matrix $c_{ij}$ between nodes
$i \in V_1$ and $j \in V_2$.\\
A spatial \textbf{utility function} $u_{ij}$ defining how agents
associated with node $i$ evaluate node $j$.

\Ensure simulated bipartite allocation
$\hat{p}_{ij} \equiv p(i \mid j)$.

\State Initialize simulated demand counts $D_{ij}=0$
for all $i \in V_1$ and $j \in V_2$.

\For{$i \in V_1$}
    \State Compute the utility vector:
    \(
    \mathbf{u}_i=(u_{ij})_{j\in V_2}
    \)
    \State \parbox[t]{0.87\linewidth}{Convert utilities into choice probabilities via softmax (temperature $\tau=1$): \\
    \( \displaystyle
    p_{ij}
    \equiv
    p(j\mid i)
    =
    \frac{\exp(u_{ij})}
    {\sum_{j' \in V_2}\exp(u_{ij'})},
    \)
     }
    \For{$n \in \{1,\dots,N_i\}$}
        \State \parbox[t]{0.87\linewidth}{Agent $n$ associated with node $i$ selects
        $j' \in V_2$ according to
        $\mathbf{p}_i=(p_{ij})_{j\in V_2}$.
        }
        
        \State Update the simulated demand count:
        \(
        D_{ij'} \leftarrow D_{ij'}+1.
        \)
    
    \EndFor
\EndFor

\For{$j \in V_2$}
    \State Compute total demand:
    \(
    D_{\cdot j} = \sum_{i \in V_1} D_{ij}
    \)
    \For{$i \in V_1$}
        \State
        \(
        \hat{p}_{ij} =
        \dfrac{D_{ij}}{D_{\cdot j}}, \; \text{if } D_{\cdot j}>0, \; otherwise \; 0
        \)
    \EndFor
\EndFor

\State \Return $\mathbf{\hat{p}}$
\end{algorithmic}
\end{algorithm}

Spatial-choice models are inspired by spatial interaction and marketing models~\cite{Huff1963}. In these models, nodes in $V_2$ attract demand according to their \textit{attractiveness}, while transaction costs between nodes in $V_1$ and $V_2$ act as friction. Agents associated with nodes in $V_1$ independently choose among nodes in $V_2$, so the demand generated by one node in $V_1$ does not affect the availability faced by the other nodes. Therefore, the spatial-choice models represent a non-competitive allocation mechanism.

We visualize the size of each node in $V_1$ as a population of demand agents. Each node generates demand, and its agents choose among nodes in $V_2$ according to a spatial utility function. The agents associated with each node in $V_1$ choose independently, with fixed utilities defined as a function of attractiveness and transportation cost. Hence, the demand generated by one node does not affect the availability, price, or attractiveness faced by the others.
The resulting demand is aggregated and normalized across $V_1$ to obtain the simulated bipartite allocation $\hat{p}_{ij}$. Algorithm~\ref{alg:spatial_ABM} presents the overall ABM formulation and computational steps.

This population-agent-based approach is mathematically flexible and utility-agnostic. The choice probabilities are modeled with a temperature-scaled softmax formulation, which is equivalent to a multinomial logit model~\cite{sutton2018:RL,bishop2006:ML,mcfadden1974:conditional}.
This approach can transform any utility function into choice probabilities, allowing different utility specifications to be tested. Moreover, it can also handle negative utilities because the softmax passes the score through an exponential function, which converts them into positive choice probabilities.

The agents' independent probabilistic choices (Steps 9--12 in Algorithm~\ref{alg:spatial_ABM}) can alternatively be implemented in a deterministic mean-field form by replacing individual draws with the expected aggregate demand, $D_{ij}=N_i p(j\mid i)$. This formulation is valid when agents' utilities are independent of other agents' decisions. In our empirical Huff implementation, we adopt this mean-field formulation to obtain the population-weighted allocation probabilities $\hat{p}_{ij}=p(i\mid j)$. We then generate simulated observations by sampling the distribution of each amphora type across consumption sites from a multinomial distribution, keeping its total recorded frequency fixed. 

\subsubsection{Huff gravity}
\label{subsubsec:huff}

The Huff model~\cite{Huff1963} is a probabilistic gravity-like model originally developed to describe destination choice in a spatial context. In its classical form, the probability that an agent $i$ selects store $j$ is given by:
\(
\displaystyle
p_{ij} =
\frac{A_j^{\lambda} D_{ij}^{-\beta}}
{\sum_k A_k^{\lambda} D_{ik}^{-\beta}}
\)
where $A_j$ denotes the attractiveness of store $j$. Attractiveness is often associated with the size of
the store, product variety, economic activity, or reputation. $D_{ij}$ is the distance or travel cost
between agent $i$ and store $j$. The parameters $\lambda$ and $\beta$ control the strength of
attractiveness and distance deterrence, respectively. Normalizing across all possible alternatives ensures that the probabilities sum to one.

For the archaeological case study, the Huff model can be expressed equivalently through a utility formulation, which is transformed into choice probabilities by the softmax. Each agent at consumption site $i$ makes one probabilistic choice among the 131 amphora types. The utility associated with amphora type $a$ is:
\begin{equation*}
u_{i,a} =
\overbrace{
\lambda \ln(\text{attractiveness}_{r(a),g(a)})
}^{\text{Unknown}}
-\beta \ln(T_{i,o(a)})
\end{equation*}
where $r(a)$ is the major production region, $g(a)$ is the amphora-borne good, and $T_{i,o(a)}$ is the ORBIS transport time from the production origin of amphora type $a$ to consumption site $i$. We assume that attractiveness depends on the combination of major production region and amphora-borne good, rather than on the individual amphora type.

Since explicit historical data on this attractiveness
are unavailable, we represent the unobserved component
through an \textit{Alternative-Specific Constant},
$ASC_{r(a),g(a)}$. The ASC is defined at the utility
scale and replaces the unknown attractiveness term.
Thus, $\exp(ASC_{r(a),g(a)})$ is the corresponding
multiplicative attractiveness factor in the Huff
choice probability. Then we have:
\begin{equation}
\label{eq:huff-utility-ASC}
u_{i,a} =
ASC_{r(a),g(a)}-\beta\ln(T_{i,o(a)})
\end{equation}
Then, the utilities are converted into choice probabilities using softmax normalization across all amphora types. The $ASC_{r(a),g(a)}$ is specific to each major production region and amphora-borne good, and shared across all consumption sites. It captures residual region--good advantages not explained by transport time alone, while leaving open whether these advantages reflect production capacity, institutional support, product preference, or other unobserved historical factors. One ASC is fixed to zero as a global reference (without implying zero attractiveness). The current dataset contains 12 represented region--good combinations, yielding 11 free ASCs.

Finally, in order to keep the number of parameters low and reduce the risk of overfitting, we use a single $\beta$ parameter shared across all regions, amphora-borne goods, and consumption sites. This parameter controls how strongly transport time affects the agents' choice probabilities.

\subsection{Results}
\label{subsec:eval-spatial-PRD}

\textbf{Experimental configuration:} We evaluate the Huff model under the population-weighted demand assumption, where the number of agents at each consumption site is proportional to its estimated population. Each agent chooses probabilistically among the 131 amphora types, using the utility function in Eq.~\ref{eq:huff-utility-ASC}. We infer 12 parameters through SBI: 11 attractiveness parameters (ASCs), corresponding to the observed major production region--good combinations with one global reference fixed to zero, and one shared transport-cost parameter $\beta$. The simulated allocations are normalized across consumption sites for each amphora type, consistently with the empirical observations. We use multinomial sampling, keeping the observed total frequency of each amphora type fixed. The computational settings for SBI, PPC, and SBC are summarized in Table~\ref{tab:sbi_parameters}.

\begin{figure*}[t]
\centering
\includegraphics[width=0.75\textwidth]{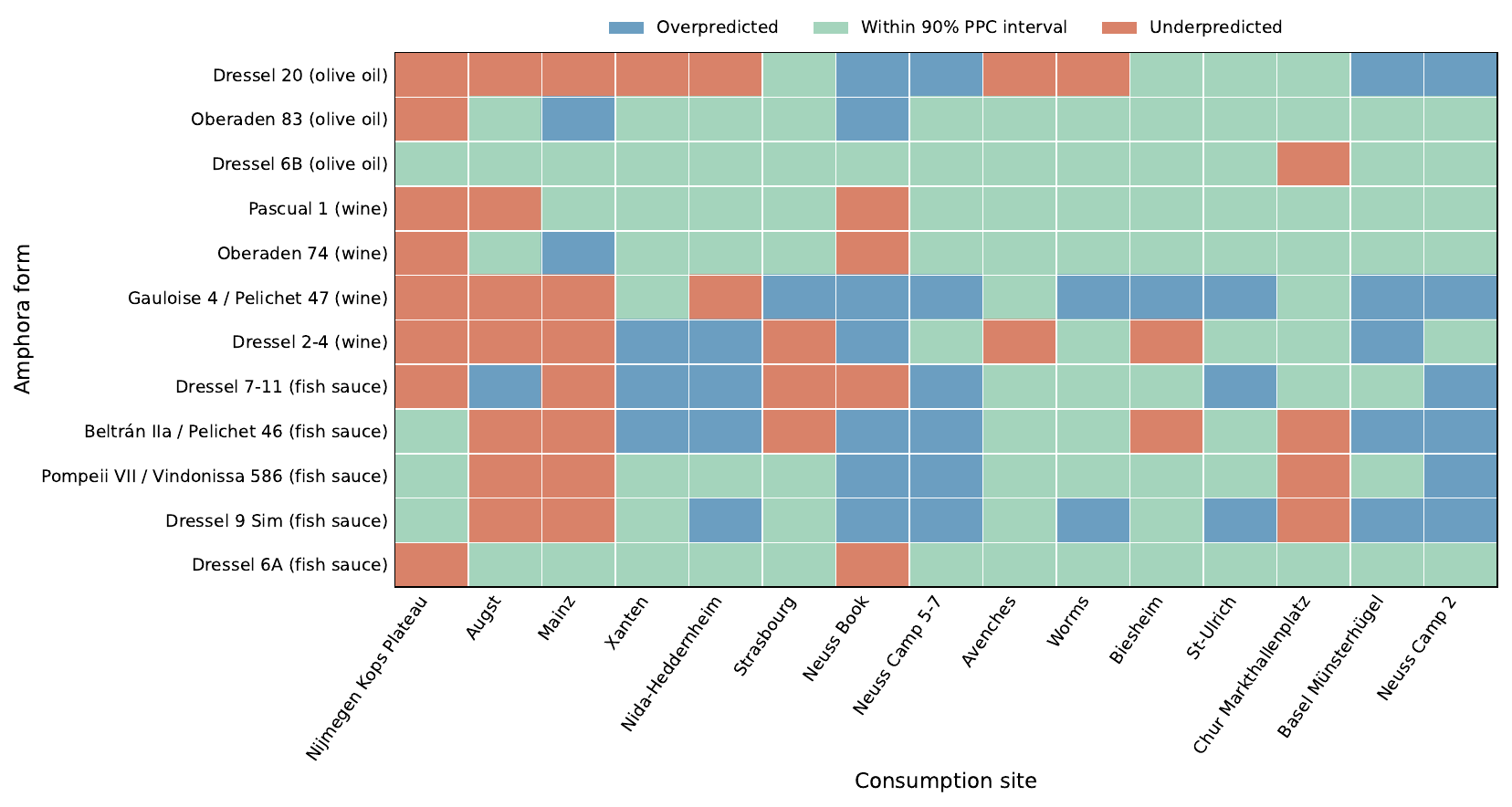}\\

\vspace{0.3cm}

\includegraphics[width=0.75\textwidth]{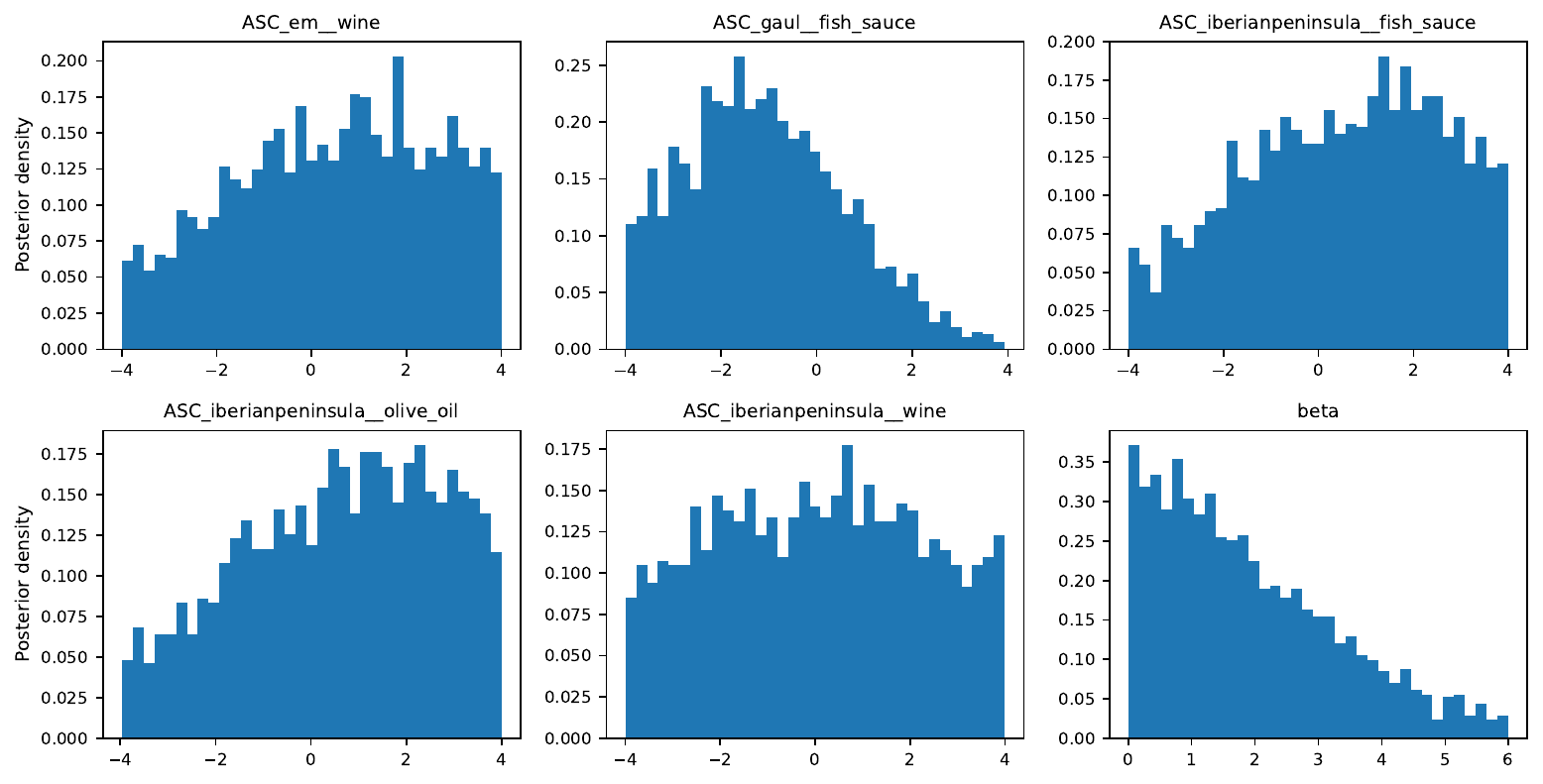}\\
\caption{Posterior predictive checks and parameter distributions for the Huff model. Top: PPC coverage across 12 selected amphora forms and 15 consumption sites. Bottom: Marginal posterior distributions of five selected region--good attractiveness parameters (ASCs) and the transport-cost coefficient $\beta$.}
\label{fig:huff_ppc}
\end{figure*}

\textbf{PPC -- model adequacy:} We evaluate the empirical adequacy of the inferred Huff model through Posterior Predictive Checks (PPC). Figure~\ref{fig:huff_ppc} summarizes the agreement between observed amphora distributions and their posterior predictive distributions.

For the PPC heatmap (Figure~\ref{fig:huff_ppc}, top), we selected 12 amphora types based on observed archaeological data, independently of model predictions. Eligible types had at least 30 recorded amphorae and occurred at a minimum of eight consumption sites. We selected three olive-oil, four wine, and five fish-sauce types, prioritizing the Iberian Peninsula and excluding Eastern Mediterranean types. We then ranked consumption sites by the number of selected amphora types recorded at each site, prioritizing sites with greater representation of different types. When sites had the same number of represented types, we ranked them by the total recorded frequency of those types. The 15 highest-ranked sites were retained. The resulting heatmap displays all 180 amphora--site combinations, including zero observations, using archaeological form names as labels.

Each heatmap cell indicates whether the observed probability falls within the corresponding 90\% posterior predictive interval, below its lower bound (model overprediction), or above its upper bound (model underprediction). Among the 180 selected combinations, 98 (54.4\%) fall within the predictive intervals, while 41 (22.8\%) are overpredicted and 41 (22.8\%) are underpredicted. The heatmap thus illustrates both agreement and discrepancies between the observed archaeological distributions and the model predictions. These results describe the selected subset rather than the complete dataset.

Across the full dataset, we obtain an overall coverage of 91.3\% for the 90\% posterior predictive intervals. However, this result is largely influenced by the sparsity of the archaeological bipartite network, in which 9,458 of 10,349 possible amphora--site combinations have a recorded frequency of zero. When considering only the 891 positive observations, coverage decreases to 44.0\%. These results indicate that the model cannot adequately reproduce many of the observed amphora distributions, consistent with the limitations discussed in the experimental rationale. The posterior parameter distributions, also shown in Figure~\ref{fig:huff_ppc} (bottom), are generally broad, suggesting substantial uncertainty in the inferred attractiveness and transport-cost parameters.

\begin{table}[t]
\label{tab:sbi_parameters}
\centering
\caption{Configuration parameters for the simulation-based inference and validation pipeline.}
\resizebox{\columnwidth}{!}{%
\begin{tabular}{lcl}
\hline
\textbf{Parameter} & \textbf{Value} & \textbf{Description} \\
\hline
Number of simulations & 5,000 & ABM simulations \\
Posterior samples & 2,000 & Inferred posterior draws \\
PPC predictive interval & 90\% & Predictive coverage level \\
PPC samples & 1,000 & Predictive realizations \\
SBC runs & 100 & Calibration trials \\
SBC samples per run & 200 & Posterior draws per trial \\
SBC credible interval & 90\% & Calibration coverage level \\
Backbone draws & 1,000 & Predictive network realizations \\
Top-$K$ edges & 50 & Edges for method comparison \\
Random seed & 0 & Reproducibility \\
\hline
\end{tabular}
}
\end{table}

\begin{figure}[t]
\centering
\includegraphics[width=0.9\columnwidth]{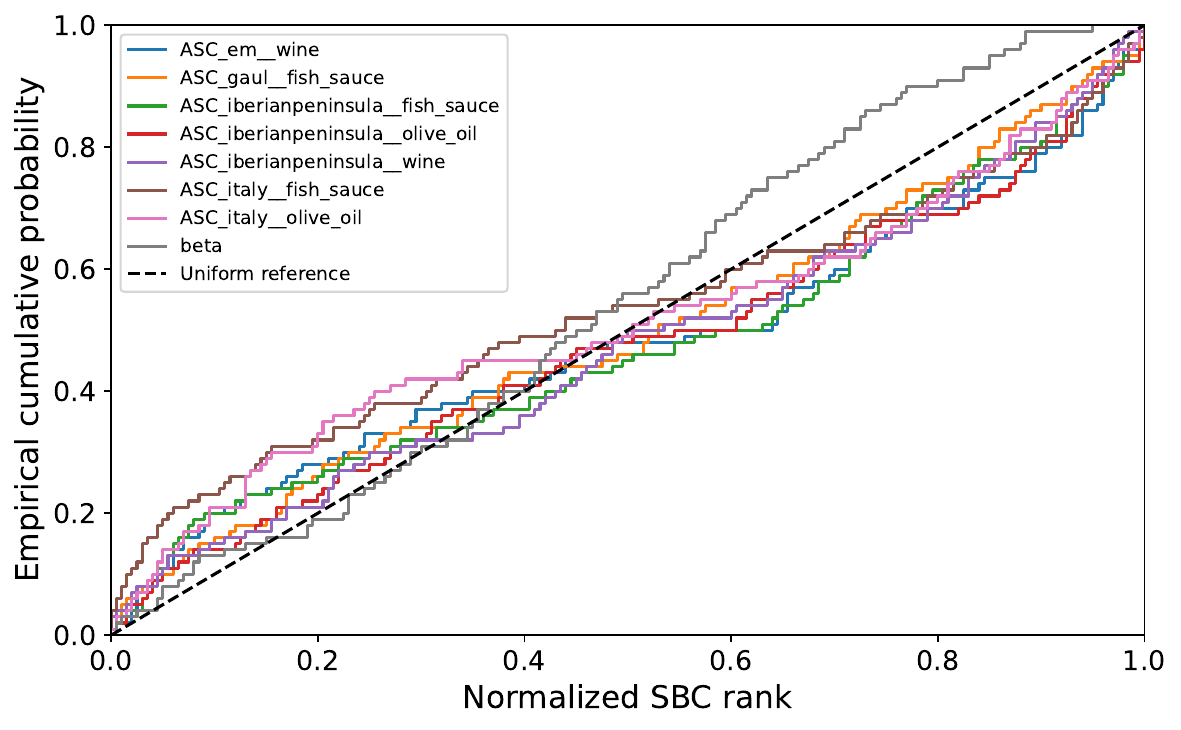}
\caption{Simulation-based calibration (SBC) of the Huff model. The colored curves show the empirical cumulative distributions of normalized SBC ranks for seven selected attractiveness parameters (ASCs) and the transport-cost parameter $\beta$, based on 100 simulation runs. The dashed diagonal represents the expected uniform distribution under calibrated inference. SBC was performed for all 12 parameters, with eight shown for readability.}
\label{fig:huff_sbc}
\end{figure}

\begin{figure}[t]
\centering
\includegraphics[width=0.47\textwidth]{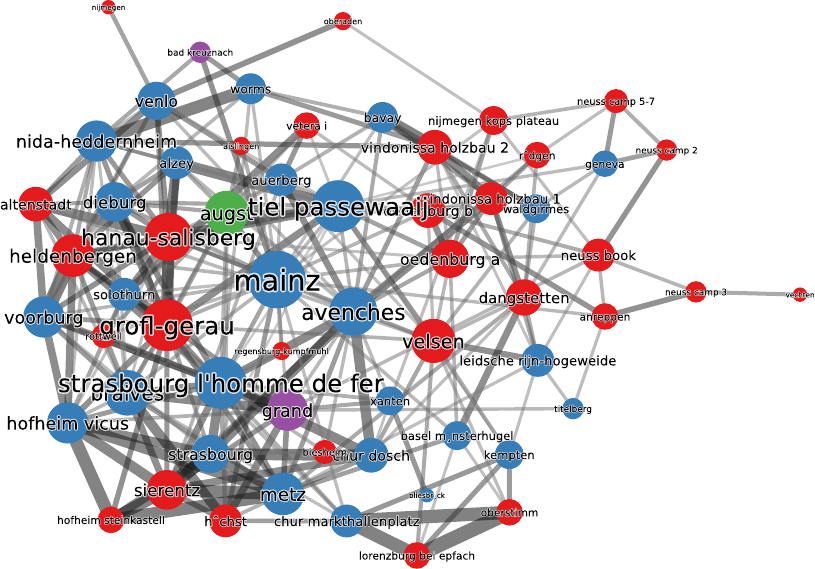}

\vspace{0.4cm} 

\includegraphics[width=0.47\textwidth]{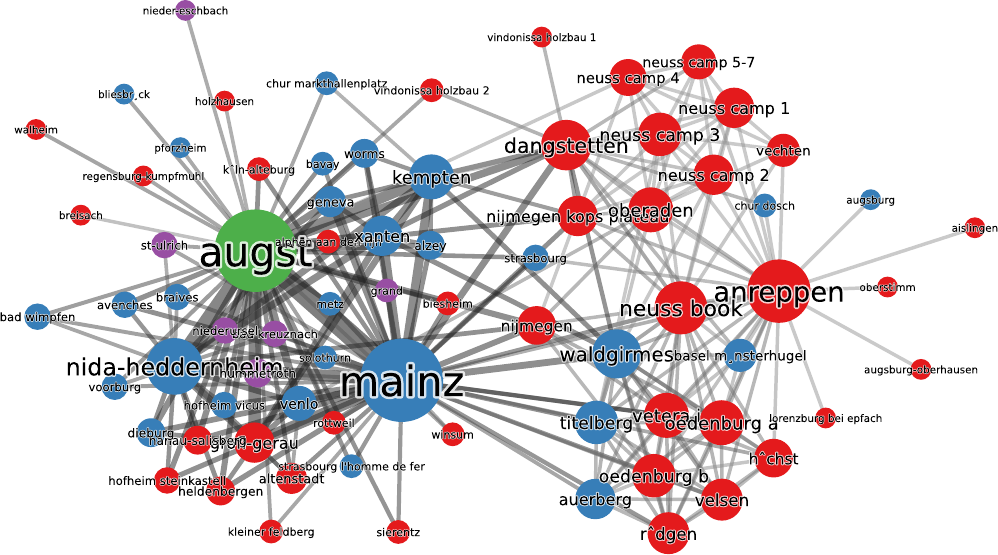}
\caption{Backbones: TDBB (top) vs NC (bottom)}
\label{fig:backbones}
\end{figure}

\textbf{SBC -- inference reliability:} We evaluate the reliability of the neural posterior estimator through Simulation-Based Calibration (SBC), using 100 simulation runs and 200 posterior samples per run. Figure~\ref{fig:huff_sbc} shows that the empirical rank distributions generally follow the expected uniform reference, although several parameters exhibit deviations from ideal calibration. These results suggest some limitations in the calibration of the neural posterior estimator. Moreover, the broad posterior distributions of the attractiveness parameters (Figure~\ref{fig:huff_ppc}) indicate substantial uncertainty in their estimation. 

\textbf{Backbone — exploratory outcome:} We construct an exploratory site--site backbone using cosine similarity between positive Pearson residuals to identify similarities that are not explained by the Huff model. We retain only positive residuals because negative residuals, particularly those associated with zero recorded frequencies, could contribute to similarity scores through shared non-observations. In this way, we focus on amphora types that occur more frequently than the Huff model predicts in a given site--site edge, rather than on similarities arising from shared zeros. For each pair of sites, we calculate a residual similarity score from the archaeological amphora frequencies and compare it with the scores obtained from posterior predictive simulations. We select the 50 most unexpected site pairs for further examination. We also compare our 50 selected site pairs with those obtained from three backbone methods: the naive weighted projection, the Noise-Corrected method, and the Disparity Filter. 

We find that only one site--site edge is shared across the methods, indicating substantial differences between the resulting backbones. These differences illustrate how the ABM assumptions influence the reconstruction of the site--site network. The unexpected similarities identified by our approach can serve as a form of theory-driven ``anomaly detection'', highlighting site pairs that warrant further archaeological investigation based on independent historical evidence.

A visual comparison of the backbones extracted using the noise-corrected (NC) baseline and our theory-driven approach (TDBB) reveals substantial differences in their network organization (Figure~\ref{fig:backbones}). The NC backbone exhibits two dominant clusters and a pronounced concentration of connectivity around Mainz and Augst, each connected to at least 49 other settlements. The NC backbone offers a valuable exploratory representation of the observed amphora distributions, but its interpretation must account for uneven archaeological documentation. Mainz and Augst are particularly well represented in our subset of the RAAD database, with 3,781 and 4,786 recorded amphorae, respectively, reflecting substantial recording and publication activity. Although both were important Roman settlements (Mainz was a provincial capital and Augst a colonia), other major centers, notably Trier and Cologne, are comparatively poorly represented in this subset of the database. Consequently, interpreting highly represented archaeological relationships as equally strong historical ties risks overlooking the effects of recording bias. 

In contrast, the TDBB backbone obtained using the Huff-based model exhibits a more pronounced core–periphery structure and a less extreme degree distribution. Although Mainz remains the most connected settlement, its degree decreases to 26 in the displayed backbone. The core of strong connections also extends beyond Mainz and Augst, encompassing settlements along the Mainz–Augst axis, spatially proximate sites around Mainz, and settlements farther north. 

On visual inspection, the TDBB backbone appears to emphasize certain relationships between geographically proximate settlements more strongly than the NC backbone, while retaining prominent long-distance connections. Strongly connected triangles are also visible in the Metz–Augst–Mainz area. Moreover, TDBB identifies a strongly interconnected group of three settlements whose mutual connections are entirely absent from the NC backbone, highlighting differences in the relationships retained by the two approaches. Importantly, these structural differences should not be interpreted as evidence that either backbone more accurately represents the historical network. Rather, they illustrate how the choice of the underlying generative assumptions influences which connections are considered informative. 

This distinction is particularly relevant in archaeological network research, where the transition from observed distributions of material remains to interpretations of past relationships is often undertaken implicitly and expressed through archaeological narratives. TDBB provides a dedicated methodological framework for making this interpretive step explicit, enabling an empirically informed exploration of the relational implications of archaeological theories. While the NC baseline evaluates connections against a fixed statistical null model, TDBB allows the underlying generative mechanism to be explicitly specified and replaced, enabling the exploration of alternative network structures under different theoretical hypotheses. This flexibility is a central advantage of the proposed framework, as it makes the interpretation of the extracted backbone conditional on an explicit and testable model of the processes generating the observed data.

\section{Conclusion}
In this paper we introduced a new approach to the network backboning problem: theory-driven backbones. While all backboning approaches currently are theory-agnostic and use generic algorithms to extract a backbone, we are the first to define a method to extract backbones informed by a theory about the edge generation process of a given network. We apply this specifically to the projection and backboning of bipartite networks. In synthetic experiments, we show that our approach is better able to reconstruct the planted backbone than the most popular backboning alternatives in the state of the art. In our case study, we show that, even if the model behind the backbone pipeline might not be able to perfectly reconstruct the original data, it is nevertheless able to provide interpretable and meaningful backbones.

Our work paves the way for a number of future developments. The most natural development is to provide a fully general theory-driven backbone process that can work directly on unipartite networks, rather than bundling the network projection process into it. A second future development concerns the computational efficiency of the simulation step: more research is needed to provide scalable inference to be able to handle networks with thousands or millions of edges. Finally, we can provide more case studies, showing the versatility of our approach, as well as doing an even deeper dive into the archaeology case study, to find a better model to fit the data.

\section*{Acknowledgment}
This work was supported by a research grant (VIL57402) from VILLUM FONDEN.

\bibliographystyle{IEEEtran}
\bibliography{references}

\end{document}